\documentclass[12pt, preprint]{aastex}

\slugcomment{Accepted for publication in {\it The Astrophysical Journal}}

\shorttitle{High-Energy Emission from BL Lac objects}
\shortauthors{B\"ottcher et al.}

\begin{document}

\title{Predictions of the High-Energy Emission from BL Lac objects: The Case of W Comae}

\author{M. B\"ottcher\footnote{Chandra Fellow} \footnote{Current address:
Department of Physics and Astronomy, Clippinger 339, Ohio University,
Athens, OH 45701, USA}}
\affil{Department of Physics and Astronomy, Rice University, MS 108, \\
6100 S. Man Street, Houston, TX 77005 - 1892, USA}
\email{mboett@spacsun.rice.edu}

\and

\author{R. Mukherjee}
\affil{Department of Physics and Astronomy, 
Barnard College and Columbia University,
New York, NY 10027, USA}
\email{muk@astro.columbia.edu}

\and

\author{A. Reimer}
\affil{Institut f\"ur Theoretische Physik IV,
Ruhr-Universit\"at Bochum,
D -- 44780 Bochum, Germany}
\email{afm@tp4.ruhr-uni-bochum.de}

\begin{abstract}
Spectral fitting of the radio through hard X-ray emission of 
BL~Lac objects has previously been used to predict their 
level of high-energy (GeV -- TeV) emission. In this paper, 
we point out that such spectral fitting can have very large 
uncertainties with respect to predictions of the VHE emission, 
in particular if no reliable, contemporaneous measurement of 
the GeV flux is available and the $\nu F_{\nu}$ peak (flux and
frequency) of the synchrotron component is not very precisely
known. This is demonstrated with the example of the radio-selected 
BL~Lac object W~Comae, which is currently on the source list of
the STACEE and CELESTE experiments, based on extrapolations
of the EGRET flux measured from this source, and on model 
predictions from hadronic blazar jet models. We show that the
best currently available contemporaneous optical -- X-ray spectrum
of W Comae, which shows clear evidence for the onset of the
high-energy emission component beyond $\sim 4$~keV and thus
provides a very accurate guideline for the level of hard X-ray 
SSC emission in the framework of leptonic jet models, still 
allows for a large range of possible parameters, resulting in 
drastically different $> 40$~GeV fluxes. We find that all acceptable 
leptonic-model fits to the optical -- X-ray emission of W Comae predict 
a cut-off of the high-energy emission around $\sim 100$~GeV. We suggest 
that detailed measurements and analysis of the soft X-ray variability 
of W Comae may be used to break the degeneracy in the choice of
possible fit parameters, and thus allow a more reliable prediction of
the VHE emission from this object. Using the available soft X-ray
variability measured by {\it BeppoSAX}, we predict a $> 40$~GeV flux 
from W~Comae of $\sim$~(0.4 -- 1)~$\times 10^{-10}$~photons~cm$^{-2}$~s$^{-1}$ 
with no significant emission at $E \gtrsim 100$~GeV for a leptonic jet 
model. We compare our results concerning leptonic jet models with detailed 
predictions of the hadronic Synchrotron-Proton Blazar model. This hadronic 
model predicts $> 40$~GeV fluxes very similar to those found for the leptonic 
models, but results in $> 100$~GeV emission which should be clearly 
detectable with future high-sensitivity instruments like VERITAS. 
Thus, we suggest this object as a promising target for VHE $\gamma$-ray 
and co-ordinated broadband observations to distinguish between leptonic 
and hadronic jet models for blazars.
\end{abstract}

\keywords{galaxies: active --- gamma-rays: theory --- 
BL Lacertae objects: individual (W Comae)}  

\section{\label{intro}Introduction}

After the detection of 6 high-frequency peaked BL~Lac objects (HBLs) with ground-based 
air \v Cerenkov telescope facilities, the field of extragalactic GeV -- TeV astronomy 
is currently one of the most rapidly expanding research areas in astrophysics. The 
steadily improving flux sensitivities of the new generation of air \v Cerenkov telescope 
arrays \citep{konopelko99,weekes02} and their decreasing energy thresholds, provides a 
growing potential to extend their extragalactic-source list towards intermediate and 
even low-frequency peaked BL~Lac objects (LBLs) with lower $\nu F_{\nu}$ peak frequencies 
in their broadband spectral energy distributions (SEDs). Detection of such objects at 
energies $\sim 40$ -- 100~GeV might provide an opportunity to probe the intrinsic 
high-energy cutoff of their SEDs since at those energies, $\gamma\gamma$ absorption 
due to the intergalactic infrared background is still expected to be negligible at 
redshifts of $z \lesssim 0.2$ \citep{djs02}.

Theoretical predictions of the high-energy emission of BL~Lac objects on the basis of their
emission at lower frequencies \citep{stecker96,cg02} are essential for careful planning of 
future observations by ground-based VHE $\gamma$-ray observatories. Such studies have
generally been restricted to considerations of the observed broadband spectral properties of
potential candidate sources and have mostly been based on non-simultaneous spectral measurements 
alone. In this paper, we point out that such considerations can have very large uncertainties
and ambiguities with respect to the predicted VHE emission. The importance and potential
scientific return of including detailed variability information in the modeling of HBLs
has been pointed out by \cite{ca99} who have shown the wide variety of correlated X-ray
and VHE $\gamma$-ray variability patterns which can result in time-dependent synchrotron-self-Compton 
models for blazars. In particular, they have pointed out that combined X-ray spectral and
variability information may be sufficient to predict the level of intrinsic TeV emission
in HBLs, even if no direct measurements at GeV -- TeV energies are available. Here, we will 
demonstrate that similar conclusions hold for LBLs, and investigate the example of the 
radio-selected BL Lac object W~Comae (= ON~231 = 1219+285; z = 0.102), which is currently 
on the source list of the STACEE and CELESTE experiments. Its GeV -- TeV source candidacy 
is based on the fact that W~Comae has been detected by the EGRET instrument on board the 
{\it Compton Gamma-Ray Observatory} at energies above 100~MeV, exhibiting a very hard 
spectrum \citep{vm95,sreekumar96}. A power-law extrapolation of the average EGRET 
0.1 -- 10~GeV flux into the multi-GeV -- TeV range yields a VHE flux well above the 
current detection threshold of both STACEE and CELESTE (see Fig. \ref{sscgraph}). 
\cite{db01} also report on the detection of a 27.3~GeV photon from this source by 
EGRET in April 1993. Furthermore, \cite{mannheim96} has predicted a TeV flux near 
the detection limit of the Whipple air \v Cerenkov telescope at the time, based on 
a proton-blazar model fit to a non-simultaneous broadband spectrum of W~Comae. 

The source was observed in a multiwavelength campaign in February 1996, covering the
electromagnetic spectrum from GHz radio frequencies to TeV energies \citep{maisack97}.
No TeV emission was detected by either Whipple or HEGRA.

While W~Comae is generally observed to exhibit a typical one-sided jet morphology
in VLBI images, \cite{massaro01} report the detection of a weak apparent counter-jet
component in 1999.13, if the brightest jet component with the flattest radio spectrum 
is identified with the core. Such a feature has not been found in any previous or
later radio maps of the source. \cite{massaro01} demonstrate that it is implausible
that this component is actually the emission from the counter-jet. Alternatively,
they suggest, e.g., that it could be due to a small-angle displacement of the jet
direction in a general configuration in which the jet is directed at a very small
average angle with respect to the line of sight.

The most detailed currently available simultaneous broadband spectrum of W~Comae has been 
measured in May 1998 \citep{tagliaferri00} and is shown in Figs. \ref{sscgraph} and 
\ref{blrgraph}. The X-ray spectrum has been measured by {\it BeppoSAX} and shows clear 
evidence for the intersection of the low-frequency (synchrotron) component and the 
high-frequency (Compton) component of the SED of W~Comae at $\sim 4$~keV. There was 
clear evidence for variability on a $\sim 10$~hr time scale in the LECS count rate 
at photon energies of 0.1 -- 4~keV, while no evidence for variability was found in 
the MECS count rate at 4 -- 10~keV and the PDS count rate at 12 -- 100~keV. A 3~$\sigma$ 
upper limit of 40~\% on the short-term variability amplitude in the MECS count rate 
could be derived in the May 1998 observations \citep{tagliaferri00}. The 4 individual 
spectral points in the EGRET energy range have been measured in March 1998, and are 
not strictly simultaneous to the {\it BeppoSAX} observations. The EGRET detection
was at low significance (2.7~$\sigma$), and allowed only a rather crude source localization
to within 1.5$^o$. We have calculated the spectrum of the source using 4 broad energy
bins, as described more fully in \S \ref{egret}.

The remainder of this paper is organized as follows: The re-analysis of the available EGRET
data is presented in \S \ref{egret}. In \S \ref{models} we describe the leptonic jet model 
which we use to reproduce the broadband spectrum of W~Comae. The modeling results and the 
model-dependent predictions for VHE emission from W~Comae will be presented in \S \ref{vhe}.
In \S \ref{variability} we discuss, how a detailed measurement and analysis of the photon-energy
dependent fast X-ray variability of W~Comae and other BL~Lac objects might be used to break
the degeneracy of model parameters still present in the pure spectral modeling using a leptonic
jet model. A comparison to the modeling results and predictions of a hadronic jet model are
presented in \S \ref{hadrons}. We summarize in \S \ref{summary}.

\section{\label{egret}Gamma-Ray Observations}

ON~231 is the suggested identification of the EGRET source 3EG~J1222+2841, and the
association of the EGRET source with this BL~Lac object is based on probabilistic
arguments. 3EG~J1222+2841 was a weak EGRET source that was never detected at
$> 6 \; \sigma$ in any individual viewing period (VPs). The locations of 
low-confidence EGRET sources are not well-determined owing to the wide point spread
function (PSF) at photon energies $> 100$~MeV \citep{thompson93,mhr01}, and identification
of EGRET sources with counterparts based on position alone is difficult. Analysis of
EGRET data from 1991 -- 1995 yielded a 7.7~$\sigma$ detection and a hard spectrum
with photon index $\alpha = 1.73 \pm 0.18$ \citep{hartman99}, and 3EG~J1222+2841
was identified with ON~231 with ``high confidence''. However, this identification
was based on the $E > 1$~GeV position of \cite{lm97} rather than the $E > 100$~MeV
position, as is standard practice for EGRET sources.

\cite{mhr01} have recently done a quantitative re-evaluation of potential radio
identifications for the 3EG radio sources by calculating the probability of each
identification. They note that based on its $E > 100$~MeV position, 3EG~J1222+2841
cannot be included in their list of high confidence identifications. Using the 
$E > 1$~GeV position, \cite{mhr01} get the probability of identification to be 
4~\%, which is classified as ``plausible''. Future observations with GLAST
will be important in determining the source position with more accuracy, and
securing a more confident identification. For the purpose of our analysis,
we assume that W~Comae is the counterpart of 3EG~J1222+2841. However, the
low significance of the $\sim 2.7 \, \sigma$ detection in May 1998 and its
non-simultaneity to the BeppoSAX observations prevents us from deriving any 
strong constraints from the EGRET flux or spectrum.

\begin{figure}
\plotone{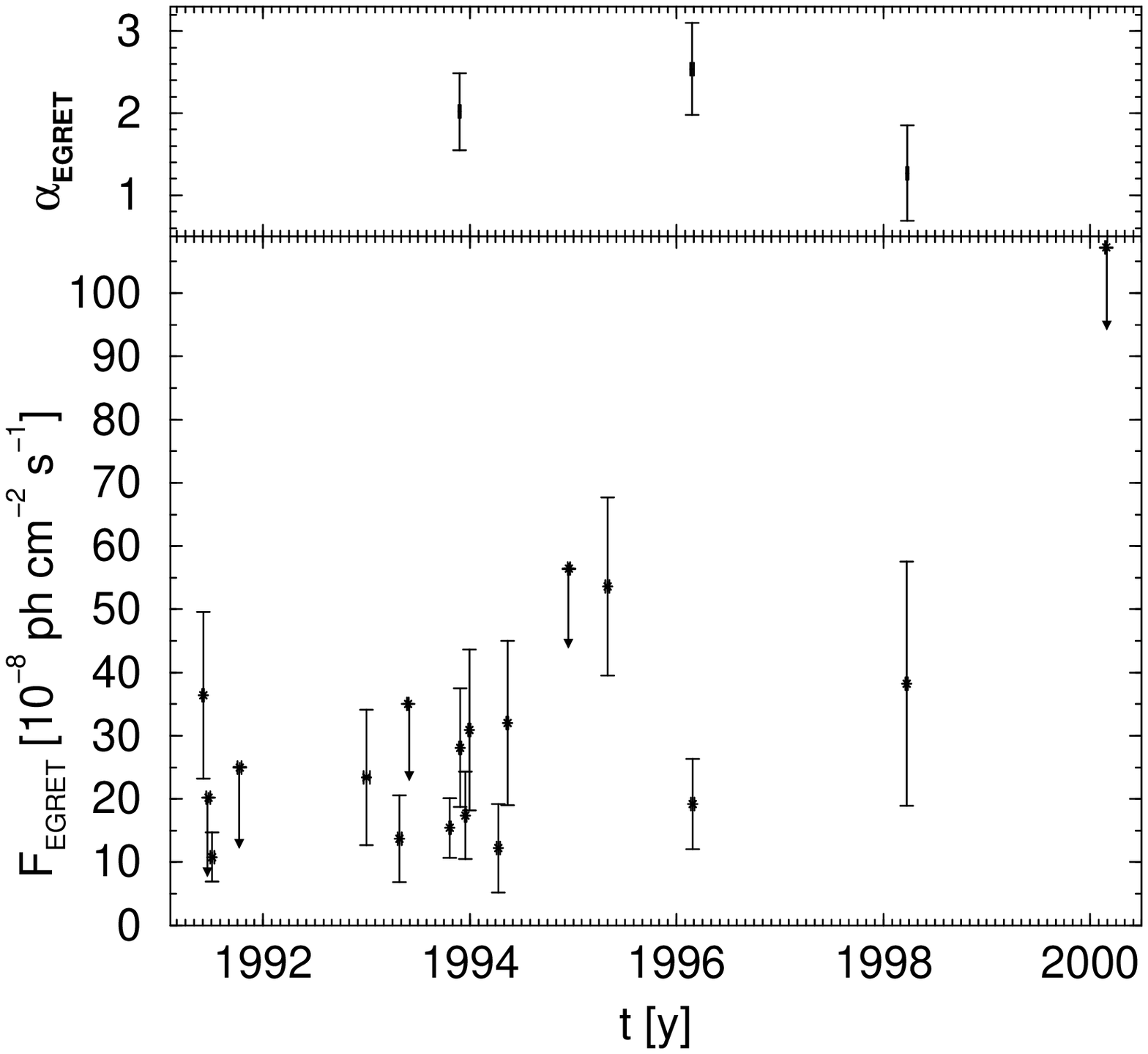}
\caption{Lower panel: EGRET light curve of W~Comae over the entire lifetime
of the {\it Compton Gamma-Ray Observatory}. Upper panel: Best-fit spectral 
indices (photon indices) of some of the most significant EGRET detections.}
\label{fluxhist}
\end{figure}

EGRET has observed 3EG~J1222+2841 = W~Comae several times since its launch 
in 1991. Table \ref{history} lists the VPs during which the source was 
observed, and the corresponding integral fluxes for energies greater than 
100~MeV. Only photons with inclination angles less than $30^\circ$ were 
used for the analysis. For Phase 1 through Cycle 4 of the EGRET observations 
(1991 -- 1995), Table \ref{history} lists data from the 3EG catalog 
\citep{hartman99}.  Three additional observations were made in Cycles 5, 7 
and 9. We have analyzed these data using the standard EGRET data processing 
technique, as described in \cite{mattox96} and \cite{hartman99}, and
included them in the table. The light curve for 3EG J1222+2841 is shown 
in Fig. \ref{fluxhist}. The figure shows fluxes for all detections at a 
level greater than $2\sigma$;  for detections below $2\sigma$, upper 
limits at the 95~\% confidence level are shown.

We have computed the background-subtracted $\gamma$-ray spectra of 
3EG~J1222+2841 for the strongest detections, as well as for March 
1998, close to the time of the {\sl BeppoSAX} observations. The 
spectra were determined by dividing the EGRET energy band of 
30 MeV -- 10 GeV into 4 bins,  and estimating the number of source 
photons in each interval, following the standard EGRET spectral
analysis technique \citep{nolan93}. We have fitted a single power law of the 
form $F(E) = k \, (E/E_0)^{-\alpha}$~photons~cm$^{-2}$~s$^{-1}$~ MeV$^{-1}$ to 
the data, where $F(E)$ is the flux, $E$ is the energy, $\alpha$ is the photon 
spectral index. Table \ref{history} includes the photon spectral indices for 
the source in the four VPs. Figures \ref{sscgraph} and \ref{blrgraph} show 
both the average EGRET spectrum (1991 -- 1995) from the 3rd EGRET catalog
\citep{hartman99} --- as dot-dashed bow-tie outline ---, as well as the 
spectrum measured in March 1998. 

In addition to the EGRET observations, W~Comae was observed by Whipple in
March -- April 1993 and March 1994. The source was not detected, and 
3~$\sigma$ upper limits of $2.3 \times 10^{-11}$~photons~cm$^{-2}$~s$^{-1}$
and $1.4 \times 10^{-11}$~photons~cm$^{-2}$~s$^{-1}$, respectively, at 
$E > 0.3$~TeV were reported \citep{kerrick95} for those two observing periods. 
Recently, the source has been of interest to the STACEE \citep{ong01} 
and CELESTE \citep{smith02} experiments, operating in the energy range 
50 -- 250~GeV. Figures \ref{sscgraph} and \ref{blrgraph} show the 
sensitivities of the two experiments. STACEE observed W~Comae in 
1998 with a prototype experiment, and reported a preliminary 
2~$\sigma$ (95~\% confidence) upper limit of $2.4 \times 
10^{-10}$~photons~cm$^{-2}$~s$^{-1}$ \citep{theoret00}, based on $\sim 5$~hr
of data. Since these results are preliminary, we have not included them
in the spectral fits shown in the figures.

\section{\label{models}Spectral modeling of W Comae using leptonic models}

For the purpose of spectral modeling using a generic leptonic jet model, it is assumed 
that a population of ultrarelativistic, non-thermal electrons and positrons is injected 
instantaneously into a spherical emitting volume of co-moving radius $R_b$. The injected
pair population is specified through a co-moving density $n_e$, low and high energy
cutoffs $\gamma_1$ and $\gamma_2$, respectively, and a spectral index $p$ so that
$n_e (\gamma) = K \gamma^{-p}$ for $\gamma_1 \le \gamma \le \gamma_2$ at the time of
injection. The location of the injection site is characterized by its height $z_i$
above the plane of a central accretion disk for which we have assumed a bolometric
luminosity of $L_D = 10^{45}$~ergs~s$^{-1}$. A magnetic field $B_0$ at the point of 
injection is chosen in equipartition with the nonthermal pair distribution at the 
time of injection, and decreases as $B = B_0 \, (z / z_i)^{-1}$. The emitting region
moves with relativistic speed $v / c = \beta_{\Gamma} = \sqrt{1 - 1 / \Gamma^2}$
along the jet which is directed at a small angle ($\theta_{\rm obs} = 1^o$) with 
respect to the line of sight. The choice of a very small observing angle implies
that the observer is located within the beaming cone of relativistically beamed
emission from the emitting region for all values of $\Gamma$ considered here, and
is consistent with the moderate superluminal motion of $\beta_{\rm app} \lesssim
2$ and the \cite{massaro01} result concerning the occasional two-sidedness 
of the radio structure observed in W~comae, as mentioned in \S \ref{intro}. 
The Doppler boosting of emission from the co-moving to the observer's frame 
is determined by the Doppler factor $D = \left[ \Gamma \, (1 - 
\beta\cos\theta_{\rm obs}) \right]^{-1}$.

Using the time-dependent radiation transfer code of \cite{bms97} and \cite{bb00},
we follow the evolution of the electron population and the radiation spectra as
the emission region moves outward along the jet. Radiation mechanisms included in
our simulations are synchrotron emission, Compton upscattering of synchrotron
photons (SSC = Synchrotron Self Compton scattering: \cite{mg85,mcg92,bm96}), 
and Compton upscattering of external photons (EC = External Compton scattering), 
including photons coming directly from the disk \citep{mk89,dsm92,ds93},as well 
as re-processed photons from the broad line region \citep{sbr94,bl95,dss97}. 
The broad line region is modelled as a spherical shell between $r_{\rm BLR, in} = 
0.2$~pc and $r_{\rm BLR, out} = 0.25$~pc, and a radial Thomson depth $\tau_{\rm T, 
BLR}$ which is considered a free parameter. 

The most rapid variability has been observed in soft X-rays, where approximately
symmetric flare profiles on a time scale of $\sim 10$~hr have been seen 
\citep{tagliaferri00}. This constrains the size of the emission region to 
be $R_b \lesssim D \times 10^{15}$~cm. For typical Doppler factors of 
$D \sim 10$, this motivates our choice of $R_b = 10^{16}$~cm, which we 
adopt throughout this paper. 

The assumption that the magnetic field is in approximate equipartition with the
emitting electron/positron population allows an independent estimate of the co-moving
magnetic field: A power-law population of electrons with spectral index $p$, emitting
a synchrotron $\nu F_{\nu}$ peak flux of $\nu F_{\nu}^{\rm pk} = 10^{-10} \, 
f_{-10}$~ergs~cm$^{-2}$~s$^{-1}$ at a dimensionless photon energy $\epsilon_{\rm pk}
= h \nu_{\rm pk} / (m_e c^2) = 10^{-6} \, \epsilon_{-6}$ requires a magnetic field of

\begin{equation}
B = 9 \, \left( {D \over 10} \right)^{-1} \, {d_{27}^{4/7} \, f_{-10}^{2/7} \, \epsilon_B^{2/7}
\over (1 + z)^{4/7} \, \epsilon_{-6}^{1/7} \, R_{15}^{6/7} \, (p - 2)^{2/7}} \; {\rm G},
\label{B_estimate}
\end{equation}
where $d_{27}$ is the luminosity distance in units of $10^{27}$~cm, $\epsilon_B$ is the
magnetic-field equipartition factor, and $R_{15}$ is $R_B$ in units of $10^{15}$~cm. For
$d_{27} = 1.45$, $f_{-10} \sim 1$, $\epsilon_B \sim 1$ (corresponding to equipartition),
$\epsilon_{-6} \sim 1$ (corresponding to a peak frequency in the near infrared), $R_{15} 
= 10$, and $p \sim 2.5$, we find $B_{\rm ep} \sim 1.8$~G, which is within a factor of
$\sim 2$ of the values quoted in Tab. \ref{parameters}. The above value of $B_{\rm ep}$
implies a synchrotron cooling time scale (in the observer's frame) of electrons 
emitting synchrotron radiation at an observed energy $E_{\rm sy} = 1 \, E_{\rm keV}$~keV
of

\begin{equation}
\tau_{\rm sy} \approx 0.12 \, \left( {B \over 1.8 \, {\rm G}} \right)^{-3/2} \,
\left( {D \over 10} \right)^{-1/2} \, E_{\rm keV}^{-1/2} \; {\rm hr}.
\label{tau_sy}
\end{equation}
which, for X-ray photon energies, is shorter than the dynamical time scale 
$R_B / (D \, c)$, in agreement with the approximately symmetric shape of the 
X-ray light curves. Since the optical -- $\gamma$-ray spectrum constitutes a
time-average over a time scale of $\tau \gtrsim 20$~hr, we expect to observe
the time-averaged emission from a strongly cooled electron population down to
synchrotron energies of

\begin{equation}
E_{\rm c} = 3.5 \times 10^{-2} \, \tau_{20}^{-2} \, \left( {B \over 
1.8 \, {\rm G}} \right)^{-3} \, \left( {D \over 10} \right)^{-1} \;
{\rm eV},
\label{E_c}
\end{equation}
which is in the infrared range. In Eq. (\ref{E_c}), $\tau_{20} = \tau / 
(20 \, {\rm hr})$. If additional electron cooling mechanisms play a 
significant role (see \S \ref{variability}), the cooling time scale 
$\tau_{\rm c}$ will obviously be shorter than $\tau_{\rm sy}$, further 
strengthening the argument for time-averaged emission observed beyond 
infrared frequencies. 

\section{\label{vhe}Spectral Modeling Results and Predictions for VHE emission}

Since the EGRET detection of W~Comae during May 1998 was of rather low
significance (see \S \ref{egret}) and not quite simultaneous to the 
{\it BeppoSAX} observation, and in view of the uncertainty of the 
source identification, we first focus on the 
simultaneous optical -- X-ray spectrum. As the simplest possible variation 
of the leptonic jet model, we attempt to model this spectrum using a strongly
SSC-dominated model in which we neglect any soft-photon input from the
broad line region (i.e., we set $\tau_{\rm T, BLR} = 0$). As a first
guess, we fix a bulk Lorentz factor of $\Gamma = 10$, yielding $D =
19.41$. We then start out by choosing an arbitrary value of $\gamma_1$, 
and try to adjust the model parameters $\gamma_2$, $p$, and $n_e$ so 
that a good fit to the optical -- X-ray spectrum of W~Comae is achieved.
We find that this is possible for values of $500 \lesssim \gamma_1
\lesssim 2000$. The curves no. 1 -- 5 in Fig. \ref{sscgraph} illustrate the 
resulting model fits for a sequence of models with $\gamma_1$ in the above range.
For substantially lower values of $\gamma_1$, our model spectra become too
flat to join the optical and X-ray spectra. The model parameters for each 
simulation are listed in Table \ref{parameters}. The last column in Table 
\ref{parameters} lists the integrated photon fluxes from those models at 
energies $E > 40$~GeV. The table indicates that different SSC model fits 
to the optical -- X-ray spectrum of W~Comae predict levels of VHE emission 
which differ by a factor of more than 10. 

In a second step, we investigate how much our results depend on the
(rather arbitrary) choice of $\Gamma = 10$ adopted above. To this aim,
we now fix an intermediate value of $\gamma_1 = 1000$ from the previous
models, and attempt to find model fits for different values of $\Gamma$.
Specifically, we repeat the fitting procedure for $\Gamma = 4$ and
$\Gamma = 15$, corresponding to $D = 7.78$, and $D = 28.05$, respectively.
Again, we do not encounter any fundamental problem in finding appropriate
model parameters to provide a good fit to the optical -- X-ray spectrum
of W~Comae. Similar to the previous series of fits, the two new model fits 
predict levels of VHE emission differing by about an order of magnitude.
The fit results are shown as curves no. 6 and 7 in Fig. \ref{sscgraph}

Fig. \ref{sscgraph} illustrates another very interesting result: Although
the different SSC model fits predict very different levels of $> 40$~GeV
emission, they all cut off around 100~GeV. Thus, no $> 100$~GeV emission
is predicted by any of our model fits. This is related to the high-energy
cut-off of the synchrotron component, which is very well constrained by
the high-quality X-ray spectrum observed by {\it BeppoSAX}. Table \ref{parameters}
indicates that there is a certain degree of anti-correlation between $\gamma_1$
and $\gamma_2$. As mentioned above, the minimum allowed value for $\gamma_1$
is rather well defined so that we do not have the freedom to choose very low
values of $\gamma_1$ in order to attempt to find acceptable fits with very
large values of $\gamma_2$ which would be able to produce substantially
higher VHE cutoffs.

As pointed out by \cite{tagliaferri00}, a pure SSC model fit to the 1998
SED of W~Comae produces a $\gamma$-ray spectrum which is incompatible with
the EGRET spectrum if the spectral index resulting from this low-significance 
detection is to be trusted. The fact that EGRET detections of the source
during other viewing periods yielded similar spectral shapes in the 0.1 -- 10~GeV
regime (\S \ref{egret}) provides some circumstantial evidence that there
might indeed be a separate high-energy emission component producing a spectral
bump centered at a peak energy of $E_{\rm pk, HE} \gtrsim 10$~GeV. In the
framework of our generic leptonic jet model, this can be plausibly related
to an external Compton component. In Fig. \ref{blrgraph}, we illustrate, how
the overall spectral shape of our model fit no. 1 (see Fig. \ref{sscgraph} and
Tab. \ref{parameters}) changes if an external Compton component due to soft
photons reprocessed in the broad line region is included in the model. The
radio -- hard X-ray emission from the different models remains virtually
invariant under the inclusion of the EC component. The EGRET spectrum can
be very well accomodated assuming a radial Thomson depth of our model BLR
of $\tau_{\rm T, BLR} \sim 3 \times 10^{-3}$. This would raise the predicted
$> 40$~GeV flux from the model by a factor of $\sim 3$ compared to the
pure SSC model with otherwise identical parameters. The sharp cut-off of
the VHE emission around $\sim 100$~GeV remains unchanged even with the
inclusion of the EC component. The low-energy cut-off of the 
external-Compton component is caused by the sharp cut-off in the
electron injection function at $\gamma_1$. This cut-off is maintained
in the evolving electron distribution throughout the simulation because 
the radiative cooling time scale for electrons of energy $\gamma \lesssim 
\gamma_1$ is much longer than the dynamical time scale for the parameters
used here.

From this analysis, we can conclude that pure spectral modeling of the
optical -- X-ray emission of W~Comae is insufficient to make reliable
predictions about the expected level of VHE emission at $> 40$~GeV from 
this object. In the next section, we illustrate, how the combination of
spectral and X-ray variability information can be used to put significantly
tighter constraints on the expected level of VHE emission --- even without 
independent information about the $\gamma$-ray emission.

\section{\label{variability}X-ray variability}

Apart from the spectral information, the {\it BeppoSAX} X-ray observations 
of May 1998 revealed variability in the soft band (0.1 -- 4~keV), dominated 
by a broad flare with a rise time of $\sim 7$~hr and a decay time of 
$\sim 10$~hr (see Fig. 2 of \cite{tagliaferri00}), corresponding to
$\sim 6$~hr and $\sim 9$~hr in the cosmological rest-frame of the source. 
In the hard band (4 -- 10~keV), no significant variability was detected. 
Because of the limited photon statistics, local spectral indices or
hardness ratios within the soft and hard bands could not be extracted 
with useful temporal resolution. However, such information might become
available through observations with the new generation of X-ray telescopes,
in particular {\it XMM-Newton} and {\it Chandra}. For this reason, we are
now investigating generic light curves and spectral variability signatures
at X-rays resulting from the different model fits to W~Comae presented in
the previous section. 

To do so, we use a code which is based on the jet radiation transfer code
of \cite{bms97} and \cite{bb00}, but accounts for time-dependent electron
acceleration and/or injection throughout the evolution of the emitting
region as it moves outward along the jet. A detailed code description as
well as a parameter study for various generic model situations will be
presented in \cite{bc02}. The numerical
approach is very similar to the one used by \cite{lk00} who had
investigated the broadband spectral variability features in a pure SSC
model for flaring blazars, but our code allows for the additional
electron cooling and photon emission from external Compton scattering
(both direct accretion-disk photons and accretion-disk emission
reprocessed in the broad-line region) and takes into account 
$\gamma\gamma$ absorption intrinsic to the source, including
the corresponding pair production. 

In order to investigate a generic spectral-variability model for W~Comae, 
we assume that the electron injection during a flare occurs at a constant
rate over one dynamical time scale $t_{\rm dyn} = R_B / c$ in the 
co-moving frame, and then proceeds at a lower rate, corresponding
to the quiescent emission outside the flaring episode. Consequently,
the parameters pertaining to the relativistic electron distributions
quoted in Table \ref{parameters} correspond to flaring injection 
luminosities of $L_{\rm inj}^{\rm fl} = 3 \times 10^{41}$~ergs~s$^{-1}$
(fit no. 1) to $L_{\rm inj}^{\rm fl} = 2.6 \times 10^{42}$~ergs~s$^{-1}$
(fit no. 5). We choose a quiescent injection luminosity of 
$L_{\rm inj}^{\rm qu} = 10^{40}$~ergs~s$^{-1}$.

Fig. \ref{phspectra} illustrates the time-dependent photon spectra
resulting from these simulations corresponding to the spectral fits
no. 1 (SSC model with the lowest SSC flux of the fits shown in Fig. \ref{sscgraph}),
5 (SSC model with the highest SSC flux of the fits shown in Fig. \ref{sscgraph}),
and 10 (complete SSC + EC model with the largest EC contribution of the fits
shown in Fig. \ref{blrgraph}). Fig. \ref{lightcurves} shows the light curves
in the R-band and at 4 different X-ray energies resulting from the same time-dependent
simulations. The figure illustrates that the light curves corresponding to the
fits no. 1 and 5 are qualitatively very different. The almost symmetric shape
of the observed light curve is clearly not reproduced by the SSC dominated case
no. 5. However, no significant difference in the optical and X-ray light curves
results from the inclusion of an EC component, even in the most extreme case
of spectral fit no. 10 (with $\tau_{\rm T, BLR} = 10^{-2}$). Note, however, that
even in this case the bolometric luminosity (and, consequently, the electron cooling)
is still dominated by the synchrotron component, which may explain the fact that 
it has only a negligible impact on the low-frequency light curves.

In Fig. \ref{hic}, we show tracks in the hardness-intensity diagrams (monoenergetic
flux, normalized to the peak flux, vs. energy spectral index $\alpha$), produced in 
the different model situations. As with the light curves shown in Fig. \ref{lightcurves},
the synchrotron dominated case no. 1 should be clearly distinguishable from the SSC
dominated case no. 5, while the inclusion of an EC component leaves the hardness-intensity
tracks virtually unchanged.

In the simulations above, we have assumed the simplest possible time-dependence
of the electron acceleration, namely a step function. \cite{lk00} have also touched
on the issue of different intrinsic injection functions, and found the effect on the
light curves is only of minor importance, if the total duration of the flare injection
event is of the same order of magnitude as the dynamic time scale, and the total 
injected energy remains unchanged. From this, we may conclude that our results do
not strongly depend on the detailed shape of the injection profile, at least to the
degree of accuracy with which the soft X-ray variability could be measured by
{\it BeppoSAX}.

In summary, our modeling results indicate that cooling of the ultrarelativistic 
electron and/or pair plasma in the jet is synchrotron and/or EC dominated. The
most realistic model parameters therefore seem to be close to the simulations
no. 1 or 2, with a possible moderate contribution due to external Compton scattering,
depending on whether the observed EGRET flux from 3EG~J1222+2841 is indeed related
to W~Comae. Consequently, we predict a $> 40$~GeV flux from W~Comae of $\sim$~(0.4
-- 1)~$\times 10^{-10}$~photons~cm$^{-2}$~s$^{-1}$ with no significant emission at 
$E \gtrsim 100$~GeV. Note, however, that our conclusions about the X-ray variability
have so far only been based on the soft X-ray light curve measured by {\it BeppoSAX}.
We strongly suggest that our predictions about the energy-dependent hardness-intensity
variability illustrated in Fig. \ref{hic} should be tested with future observations
by {\it XMM-Newton} or {\it Chandra}.

\section{\label{hadrons}Comparison to hadronic models}

An alternative to leptonic models are the so-called "hadronic models" proposed to 
explain $\gamma$-ray emission from blazars. While leptonic models deal with a 
relativistic e$^\pm$ plasma in the jet, in hadronic models the relativistic jet
consists of a relativistic proton ($p$) and electron ($e^-$) component. Here we 
use the hadronic Synchrotron-Proton Blazar (SPB-) model of \cite{muecke02} to 
model the spectral energy distribution (SED) of W~Comae in May 1998.

Like in the leptonic model the emission region, or ``blob'', in an AGN
jet moves relativistically along the jet axis which is closely aligned 
with our line-of-sight. Relativistic (accelerated) protons, whose particle 
density $n_p$ follows a power law spectrum $\propto \gamma_p^{-\alpha_p}$ 
in the range $2\leq\gamma_p\leq\gamma_{\rm{p,max}}$, are injected instantaneously 
into a highly magnetized environment ($B=const$ within the emission region), 
and suffer energy losses due to proton--photon interactions (meson production 
and Bethe-Heitler pair production), synchrotron radiation and adiabatic expansion. 
The mesons produced in photonmeson interactions always decay in astrophysical
environments, however, they may suffer synchrotron losses before the decay, 
which is taken into account in this model.

The relativistic primary $e^{-}$ radiate synchrotron photons that manifest 
themselves in the blazar SED as the synchrotron hump, and serve as the target 
radiation field for proton-photon interactions and the pair-synchrotron cascade 
which subsequently develops. The SPB-model is designed for objects with a 
negligible external target photon component, and hence suitable for BL~Lac 
Objects. The cascade redistributes the photon power to lower energies where
the photons eventually escape from the emission region. The cascades can be 
initiated by photons from $\pi^0$-decay (``$\pi^0$ cascade''), electrons from 
the $\pi^\pm\to \mu^\pm\to e^\pm$ decay (``$\pi^\pm$ cascade''), $p$-synchrotron 
photons (``$p$-synchrotron cascade''), charged $\mu$-, $\pi$- and $K$-synchrotron 
photons (``$\mu^\pm$-synchrotron cascade'') and $e^\pm$ from the proton-photon 
Bethe-Heitler pair production (``Bethe-Heitler cascade'').

\cite{mp01} and \cite{muecke02} have shown that the ``$\pi^0$ cascades'' and 
``$\pi^\pm$ cascades'' generate rather featureless photon spectra, in contrast 
to ``$p$-synchrotron cascades'' and ``$\mu^\pm$-synchrotron cascades'' that 
produce a double-humped SED as typically observed for $\gamma$-ray blazars. 
The contribution from the Bethe-Heitler cascades is mostly negligible. In 
general direct proton and muon synchrotron radiation is mainly responsible 
for the high energy hump in blazars whereas the low energy hump is dominanted
by synchrotron radiation from the primary $e^-$, with a contribution of 
synchrotron radiation from secondary electrons (produced by the $p$- and 
$\mu^\pm$-synchrotron cascade). A detailed description of the model itself, 
and its implementation as a (time-independent) Monte-Carlo/numerical code, 
has been given in \cite{mp01}.

For the modeling of the 1998 SED of W Comae we have fixed the effective size
scale of the emission region to $R_b=\frac{1}{2} c t_{\rm{var}} D$ where 
$t_{\rm{var}}\approx 10$h is the measured soft X-ray variability time scale. 
We consider bulk Doppler factors in the range $D = 5$ -- 20 for the fitting 
procedure, which is consistent with the moderate superluminal motion detected 
by \cite{massaro01}. The primary relativistic electrons emit synchrotron photons, 
which serve as the target photon field for photon-proton interactions and cascading. 
The synchrotron spectrum from these electrons shows a break at around $10^{13}$~Hz  
where the synchrotron cooling time scale $t_{\rm{e,syn}} \approx 8 \times 
10^8 \, (B/1 \, {\rm G})^{-2} \, \gamma^{-1}$~sec equals the adiabatic loss 
time scale $t_{\rm{ad}} \approx \xi R_b/c$
 for a relativistic jet with 
$\xi\leq 1$, taking into account a possible non-spherical 
geometry, effects 
of a possible accelerating (i.e. non-constant speed) jet, etc.
 In the soft 
X-ray regime, the spectrum turns over steeply when the cooling time scale 
becomes shorter than the acceleration time scale $t_{\rm{acc}} = r_L / (\eta c)$
($r_L$ = Larmor radius, $\eta \leq 1$ = acceleration efficiency). Note that
the theory of plasma turbulence predicts significantly lower $\eta$ values
for electrons than for protons.

Fig.~\ref{SPBfits} shows a summary of SPB-models best representing the March/May 1998 
data. Similar to the leptonic SSC model, the SPB-model seems to be not compatible
with the March 1998 EGRET data if the May 1998 LECS + MECS + PDS data from BeppoSAX 
are modeled (see models 3 -- 5). On the other hand, the BeppoSAX LECS + MECS data 
together with the EGRET spectrum can be explained by hadronic models (see models 
1 + 2), which however underestimate the flux in the PDS energy band by about a factor 
of 3. The non-compatibility of the EGRET- with the BeppoSAX-data in the modeling
procedure might be caused by their non-simultaneity and the poor significance of the
EGRET-detection. Note also that during the May 1998 BeppoSAX observing run some 
technical problems occured which might have affected the PDS data \citep{tagliaferri00}.

In the framework of the hadronic SPB-model the hard X-ray spectrum can naturally be
explained by strong synchrotron emission from the relativistic protons provided
strong magnetic fields $B$ of several tens of Gauss exist in the emission region.
For the modeling (models 3 -- 5) we use $B = 30$ -- 40~G, $\alpha_p = 1.5$,
$\gamma_{\rm{p,max}} = 5$ -- $10 \times 10^8$, a number density ratio of 
primary, relativistic electrons to protons, $e/p\approx 0.1$, and a proton 
energy density $u_p\approx 60$ -- 150~ergs~cm$^{-3}$, somewhat above the 
equipartition value, which is not surprising during activity in the source. 
With Doppler factors $D = 12$ -- 15 the target photon energy density in the 
jet frame is $\sim 10^{10}$ -- $10^{11}$~eV~cm$^{-3}$. Because proton 
synchrotron losses dominate in the high energy region over
 pion production 
losses (see Fig. \ref{loss_timescales}), the model predicts the main 
power output at several 100~MeV due to proton synchrotron radiation, with 
a strong steepening by about 2 orders of magnitude in the GeV range, and 
$\gamma$-ray emission from the $\pi$-cascades extending up to about 100~TeV
with a break at about 10~TeV at the source. Photons above $\sim 100$~GeV 
(in the co-moving frame) will be subject to $\gamma\gamma$ absorption within 
the emission region, and initiate electromagnetic cascades in the jet (see Fig. 
\ref{tau_gg}). Absorption of multi-GeV/TeV-photons in the cosmic background 
radiation field will further alter the observed spectrum: the optical depth 
exceeds unity above 300 -- 700~GeV for W~Comae. The predicted photon flux 
above 40~GeV (see Tab. \ref{SPBparameters}) is therefore similar to the 
corresponding predictions from the leptonic models (see Tab.~\ref{parameters}).

If the intrinsic target photon density increases to $\sim 10^{11}$ -- 
$10^{12}$~eV~cm$^{-3}$, muon and pion production, and therefore also 
muon and pion synchrotron radiation, dominates over proton synchrotron 
radiation (see Fig. \ref{loss_timescales}). This might have been the 
case during the EGRET-observations (see Fig. \ref{SPBfits}: model 1 + 2, 
and Fig.~\ref{SPBfit3}). Because pions and muons possess a lower rest 
mass with respect to protons, their synchrotron emission peaks at higher 
photon energies than the protons' synchrotron radiation for the same 
particle Lorentz factors and magnetic fields. The parameters used for the 
models representing the EGRET data (models 1 + 2) are: $D = 8$ -- 10, $B = 40$~G, 
$\alpha_p = 1.5$ -- 2, $\gamma_{\rm{p,max}} = (1 - 3) \times 10^9$, $e/p \approx 
0.2$ -- 1.6 and a proton energy density $u_p \approx 250$ -- 300~ergs~cm$^{-3}$.
The main power output in the high-energy regime for models 1 + 2 lies therefore 
at $\sim 10$~GeV, and is somewhat higher at IR-energies. For these parameters,
photons beyond a few tens of GeV (in the co-moving frame) will be subject to
$\gamma\gamma$ absorption, and initiate electromagnetic cascades (see Fig.
\ref{tau_gg}). Also here, the model predicts TeV-emission from the 
$\pi$-cascades (however not extending above 1~TeV), which however will 
partly be absorbed in the cosmic photon background.

Models involving meson production inevitably predict neutrino emission due to the 
decay of charged mesons. The SPB-model for W Comae in 1998 predict a $\nu_\mu + 
\bar\nu_\mu$ output of about $10^{-7}$~GeV~s$^{-1}$~cm$^{-2}$ peaking at around 
$10^{8.5-9}$~GeV. The neutrino power at $10^6$~GeV is about 
$10^{-11 \ldots -10}$~GeV~cm$^{-2}$~s$^{-1}$. No neutrino flavor
oscillations are assumed here.

In summary, the hadronic SPB-model predicts TeV-emission on a flux level near 
the detectability capabilities of CELESTE and STACEE for W Comae, but clearly 
above the sensitivity limit of future instruments like VERITAS. While leptonic models 
predict integral fluxes at $>40$~GeV for W Comae on a similar level than hadronic 
models do, TeV-emission detectable with very high-sensitivity instruments is only 
predicted for the hadronic emission processes. This is in contrast to leptonic models, 
and may therefore be useful as a diagnostic to distinguish between the hadronic and 
leptonic nature of the high-energy emission from W~Comae, in addition to its possible 
neutrino emission.

\section{\label{summary}Summary}

We have presented detailed modeling of the best currently 
available simultaneous broadband SED of the radio-selected 
BL~Lac object W~Comae, comparing state-of-the-art leptonic 
and hadronic jet models. The richest and most detailed 
portion of the SED consists of the BeppoSAX LECS + MECS
+ PDS spectrum from $\sim 0.1$ -- 100~keV, measured in 
May 1998 \citep{tagliaferri00}. It showed the low-energy 
(synchrotron) component extending out to $\sim 4$~keV, 
exhibiting significant variability on time scales of 
$\lesssim 10$~hr, and the onset of the high-energy component
beyond $\sim 4$~keV, with no evidence for short-term
variability. The SED was supplemented by simultaneous
radio and optical flux measurements as well as a
weak EGRET detection in March 1998, i. e. about 2
months before the BeppoSAX spectrum was taken. We
have done a careful re-analysis of all available EGRET
pointings on W~Comae, and confirmed that the source
exhibited an unusually hard GeV spectrum during the
March 1998 observation, with a photon spectral index
of $\alpha = 1.27 \pm 0.58$. 

Our fits using leptonic jet models yielded the following
main results:

(1) Acceptable fits to the optical to hard X-ray spectrum 
of W~Comae are possible with a rather wide range of parameters
in a synchrotron-self-Compton (SSC) dominated model. In agreement
with earlier results of \cite{tagliaferri00}, we find that such
fits are generally inconsistent with the (not quite simultaneous)
EGRET spectrum of March 1998 as well as the average spectrum
over the entire lifetime of EGRET. 

(2) The different SSC fits to the optical -- hard X-ray
spectrum of W~Comae in May 1998 result in $> 40$~GeV fluxes
of $\sim (0.4 - 4) \times 10^{-10}$~photons~cm$^{-2}$~s$^{-1}$,
but virtually no emission beyond $\sim 100$~GeV. 

(3) Including the information contained in the X-ray variability
measured by BeppoSAX, we can narrow down the range of possible 
leptonic model parameters to predict $> 40$~GeV fluxes of
$\sim (0.4 - 1) \times 10^{-10}$~photons~cm$^{-2}$~s$^{-1}$.

(4) In order to reproduce the March 1998 EGRET spectrum 
together with the May 1998 BeppoSAX spectrum, an external Compton 
component is required. Such fits result in $> 40$~GeV fluxes 
of $\sim (0.5 - 1) \times 10^{-10}$~photons~cm$^{-2}$~s$^{-1}$,
and a strong cutoff at $\lesssim 100$~GeV, as in the case of
SSC dominated models.

Successful fits to the SED of W~Comae were also possible 
using the hadronic Synchrotron-Proton Blazar model, yielding
the following main results:

(5) A model with proton-synchrotron dominated hard X-ray
to GeV $\gamma$-ray emission is well suited to reproduce 
the entire radio -- hard X-ray spectrum of W~Comae in
May 1998. Just as the SSC-dominated leptonic models, it
is inconsistent with both the March 1998 and the average
EGRET spectrum from the source. 

(6) The hadronic fit to the radio -- hard X-ray spectrum
of W~Comae predicts a $> 40$~GeV flux of $\sim (0.7 -
1.4) \times 10^{-10}$~photons~cm$^{-2}$~s$^{-1}$, i.e.
of the same order as the predictions of the leptonic jet
models. However, in contrast to the leptonic models, the
high-energy emission is expected to extend beyond 1~TeV 
at a flux level of $\Phi_{> 1 \, {\rm TeV}} \sim (3 - 18) 
\times 10^{-14}$~photons~cm$^{-2}$~s$^{-1}$. Such a flux
level is well within the reach of future high-sensitivity
instruments like VERITAS.

(7) SPB models consistent with the March 1998 EGRET
spectrum under-predict the May 1998 BeppoSAX PDS hard 
X-ray spectrum. They result in higher $> 40$~GeV fluxes 
of $\sim (5 - 13) \times 10^{-10}$~photons~cm$^{-2}$~s$^{-1}$, 
but weaker TeV fluxes of $\Phi_{> 1 \, {\rm TeV}} \sim 
(0.7 - 9) \times 10^{-14}$~photons~cm$^{-2}$~s$^{-1}$. 
In this case, STACEE and CELESTE may be able to get a weak 
detection of the source, and it would still be a promising 
candidate for detection by the future VERITAS array.

In conclusion, leptonic and hadronic jet model fits to
W~Comae make drastically different predictions with
respect to the expected very-high energy emission beyond
$\sim 100$~GeV. A detection of W~Comae at those photon
energies with future, high-sensitivity air \v Cerenkov
detector arrays would pose a serious challenge to leptonic 
jet models, and might favor hadronic models instead.

\acknowledgments
We thank D. Smith for inspiring discussions, and G. Ghisellini 
for making the data on the May 1998 SED of W~Comae available to
us. We are also grateful to J. Chiang for careful reading of the
manuscript and helpful comments. The work of MB was supported by 
NASA through Chandra Postdoctoral Fellowship grant PF~9-10007 
awarded by the Chandra X-ray Center, which is operated by the 
Smithsonian Astrophysical Observatory for NASA under contract 
NAS~8-39073. RM  acknowledges support from NSF grant PHY-9983836.
AR thanks the Bundesministerium f\"ur Bildung und Forschung for 
financial support through DESY grant Verbundforschung 05CH1PCA6.

\newpage

\newpage

\begin{figure}
\plotone{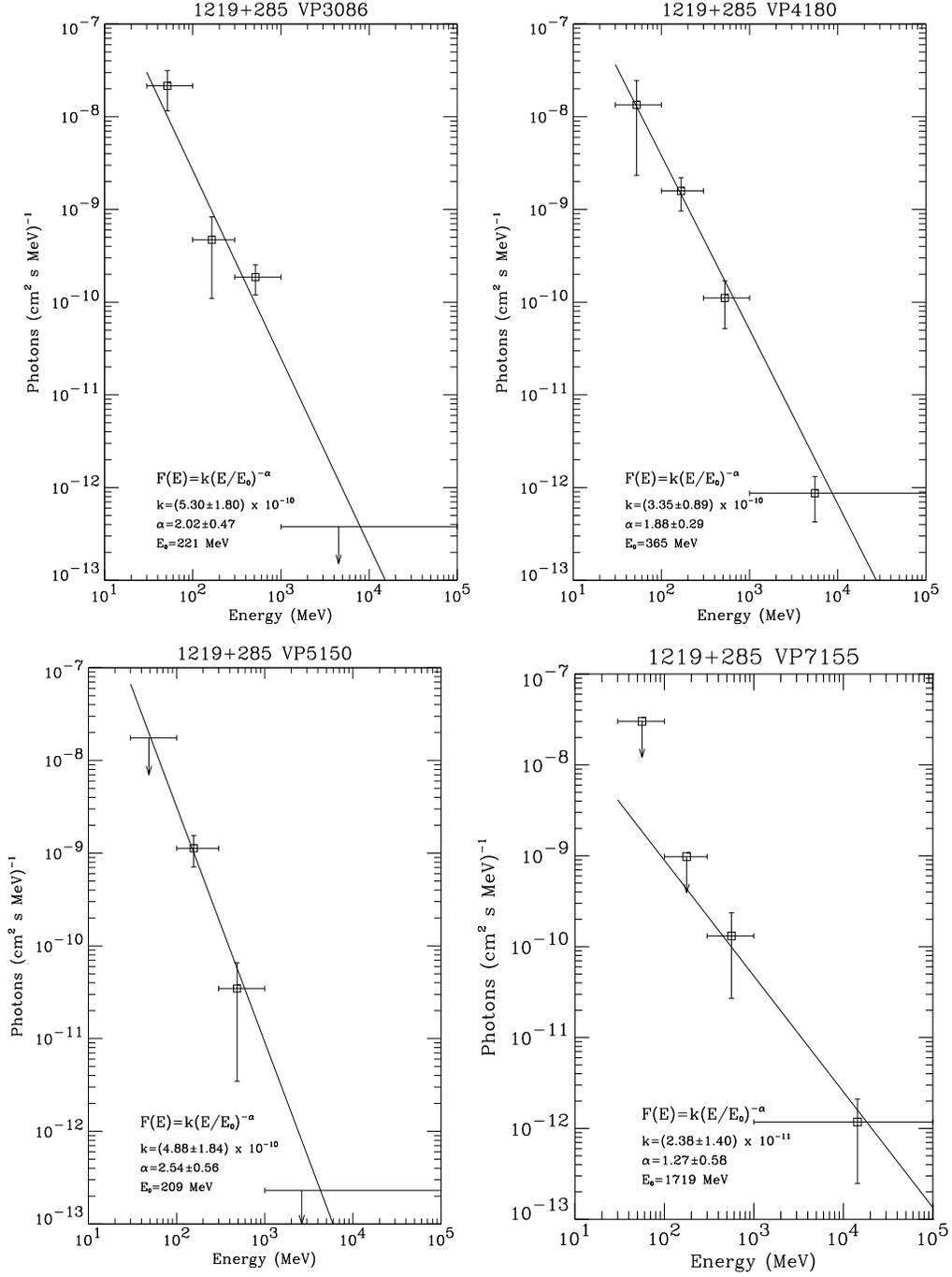}
\caption{4-point EGRET spectra during the most significant EGRET 
detections of W~Comae. The lines show the power-law fits to those 
spectra.}
\label{egretspectra}
\end{figure}

\newpage

\begin{figure}
\plotone{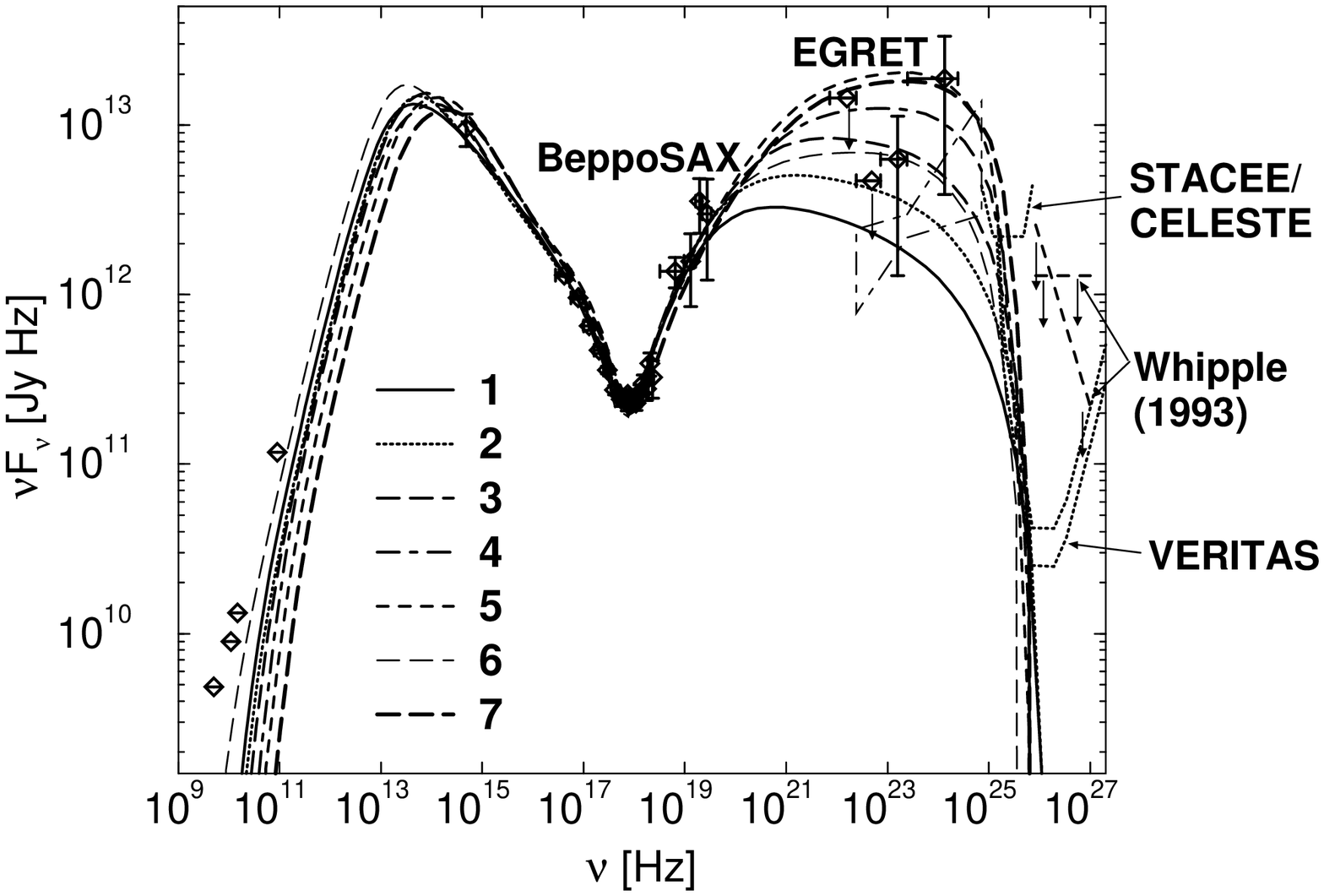}
\caption{Various model fits to the optical -- X-ray spectrum of W Comae in 
May 1998, using a pure SSC model. The radio -- X-ray spectral data points 
from \cite{tagliaferri00}; the EGRET data points are the result of our 
re-analysis of the March 1998 observation. The dot-dashed bow-tie outline 
shows the average EGRET spectrum from the 3rd EGRET catalogue \citep{hartman99}. 
Also shown are the sensitivities of current and future ACTs as well as the upper 
limit from the Whipple observation in 1993. The two different sensitivity
curves for VERITAS result from assuming an underlying power-law photon spectrum
of index $\alpha = 2.5$ and $\alpha = 3.5$, respectively. For fits no. 1 - 5, 
we have fixed the Doppler boosting factor $D = 19.41$, changed the low-energy 
cutoff of the electron distribution from $\gamma_1 = 500$ to 1600, and adjusted 
the remaining parameters to achieve an acceptable fit to the optical -- X-ray 
spectrum. For fits 6, 3, and 7, we fixed $\gamma_1 = 1000$, changed the Doppler 
boosting factor from $D = 7.78$ to $28.05$, and adjusted the remaining parameters 
to achieve a good fit. For the complete list of parameters, see Table 
\ref{parameters}.}
\label{sscgraph}
\end{figure}

\newpage

\begin{figure}
\plotone{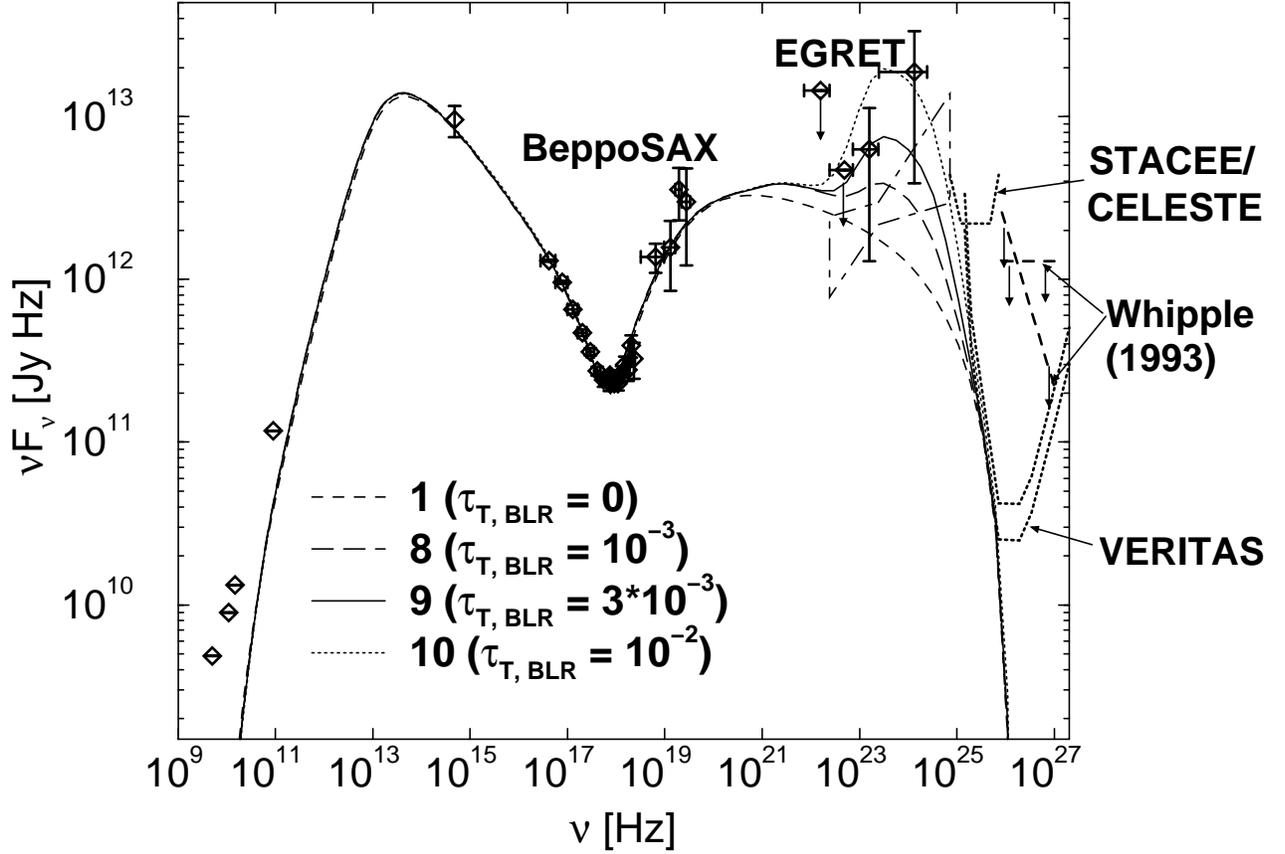}
\caption{Various model fits to the optical -- X-ray spectrum of W Comae in 1998,
using an SSC + ERC model. The choice of parameters is based on fit no. 1 in Fig. 
\ref{sscgraph}, but additionally accounting for an increasing amount of reprocessed
accretion disk radiation in a broad line region of radial Thomson depth $\tau_{\rm BLR}
= 0$ to $10^{-2}$. For the complete list of parameters, see Table \ref{parameters}.}
\label{blrgraph}
\end{figure}

\newpage

\begin{figure}
\plotone{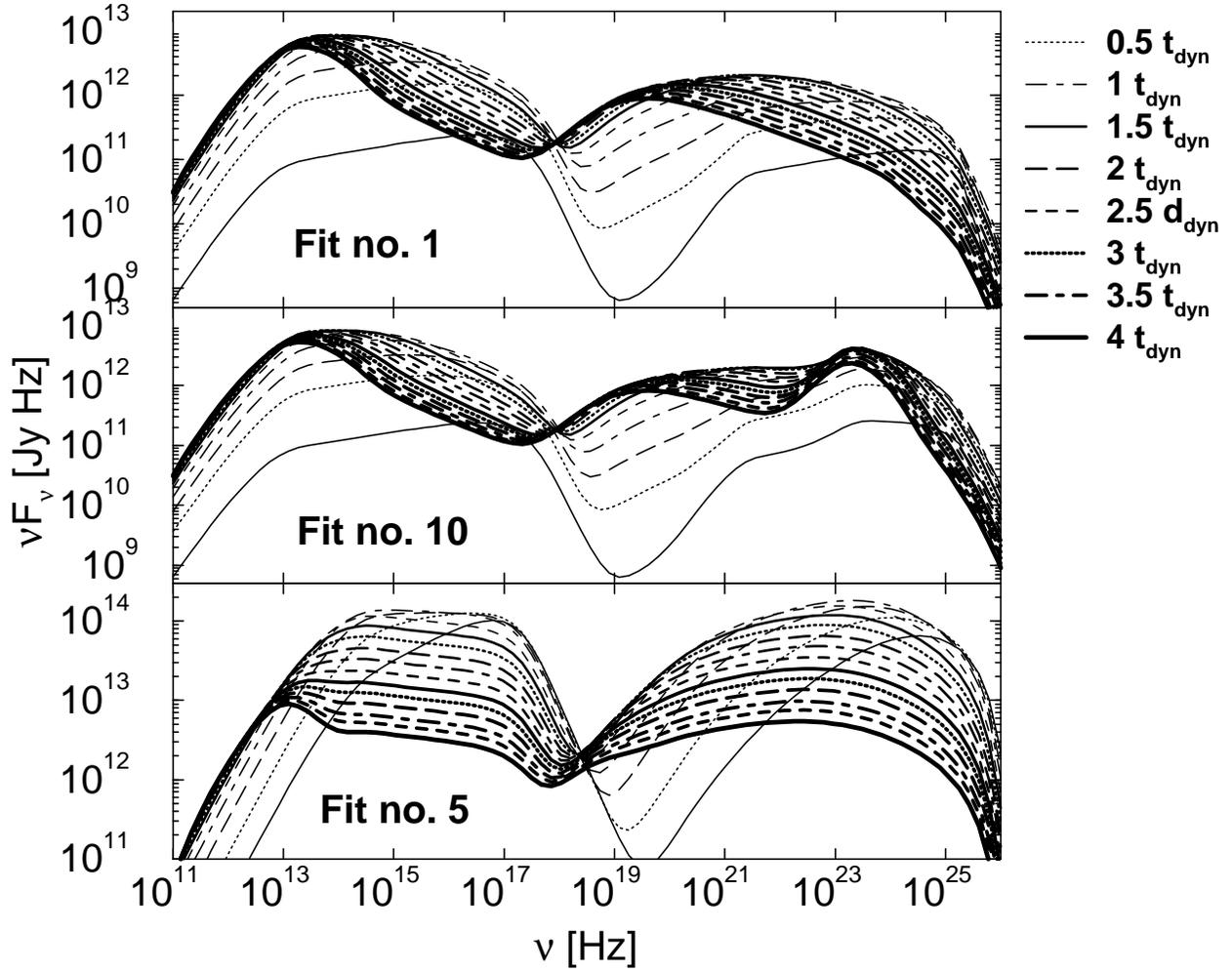}
\caption{Simulated, time-dependent broad-band spectra corresponding to the
spectral fits no. 1, 5, and 10 (see Figs. \ref{sscgraph} and \ref{blrgraph}
and Tab. \ref{parameters}). The curves are labeled by time in units of the
dynamical time scale, $R_b/(D \, c)$, which is $t_{dyn}^{\rm obs} = 1.8 
\times 10^4$~s in the observer's frame.}
\label{phspectra}
\end{figure}

\newpage

\begin{figure}
\plotone{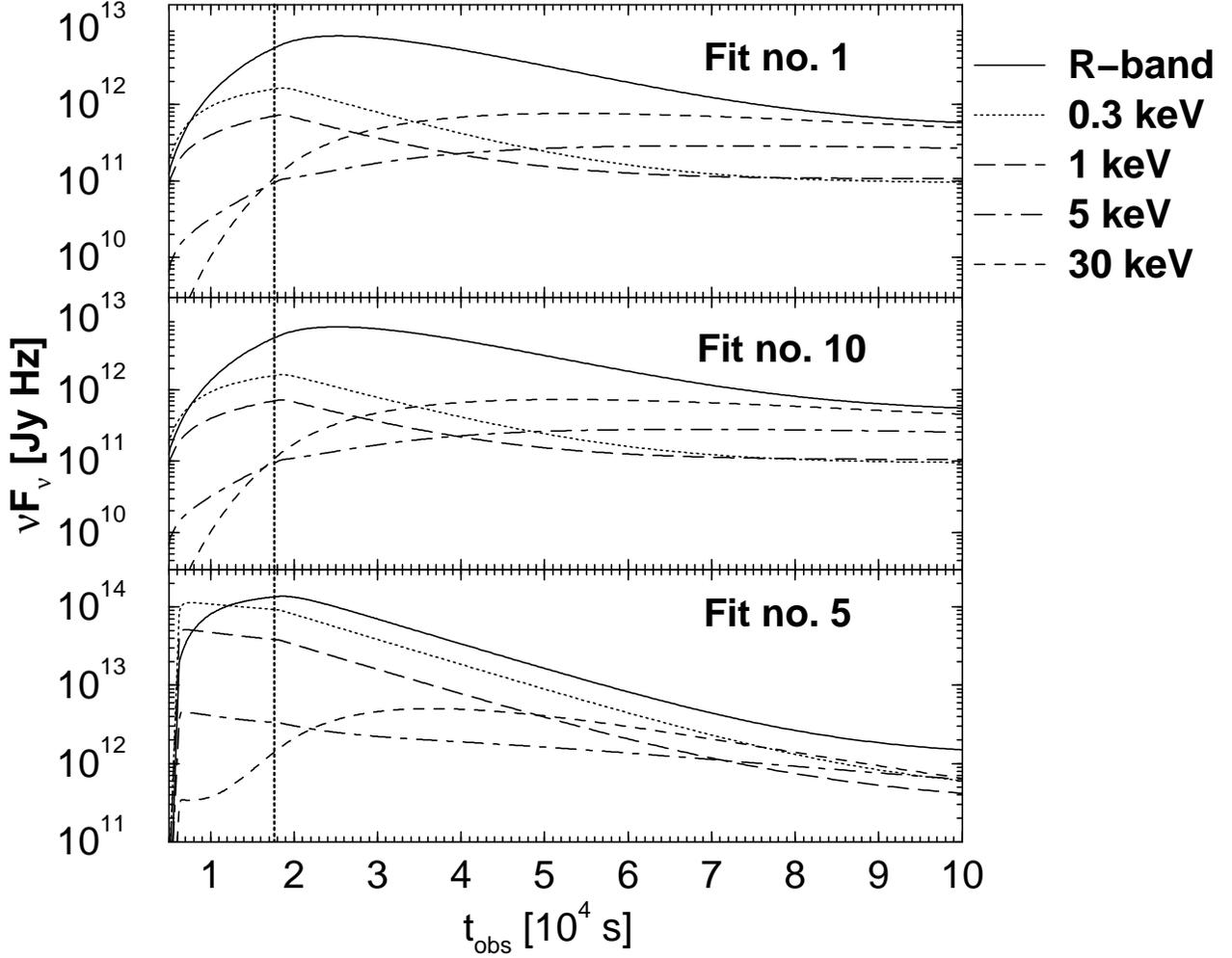}
\caption{Light curves in the R-band and at 5 different X-ray energies resulting
from the time-dependent simulations corresponding to spectral fits no. 1, 5, and 10,
illustrated in Fig. \ref{phspectra}. The dotted vertical line marks the end of
the electron-injection flare (see text for discussion). While the SSC dominated
model no. 5 produces drastically different light curves than the synchrotron
dominated cases, there is no significant difference due to the addition of a
moderate external-Compton component (fit no. 10).}
\label{lightcurves}
\end{figure}

\newpage

\begin{figure}
\plotone{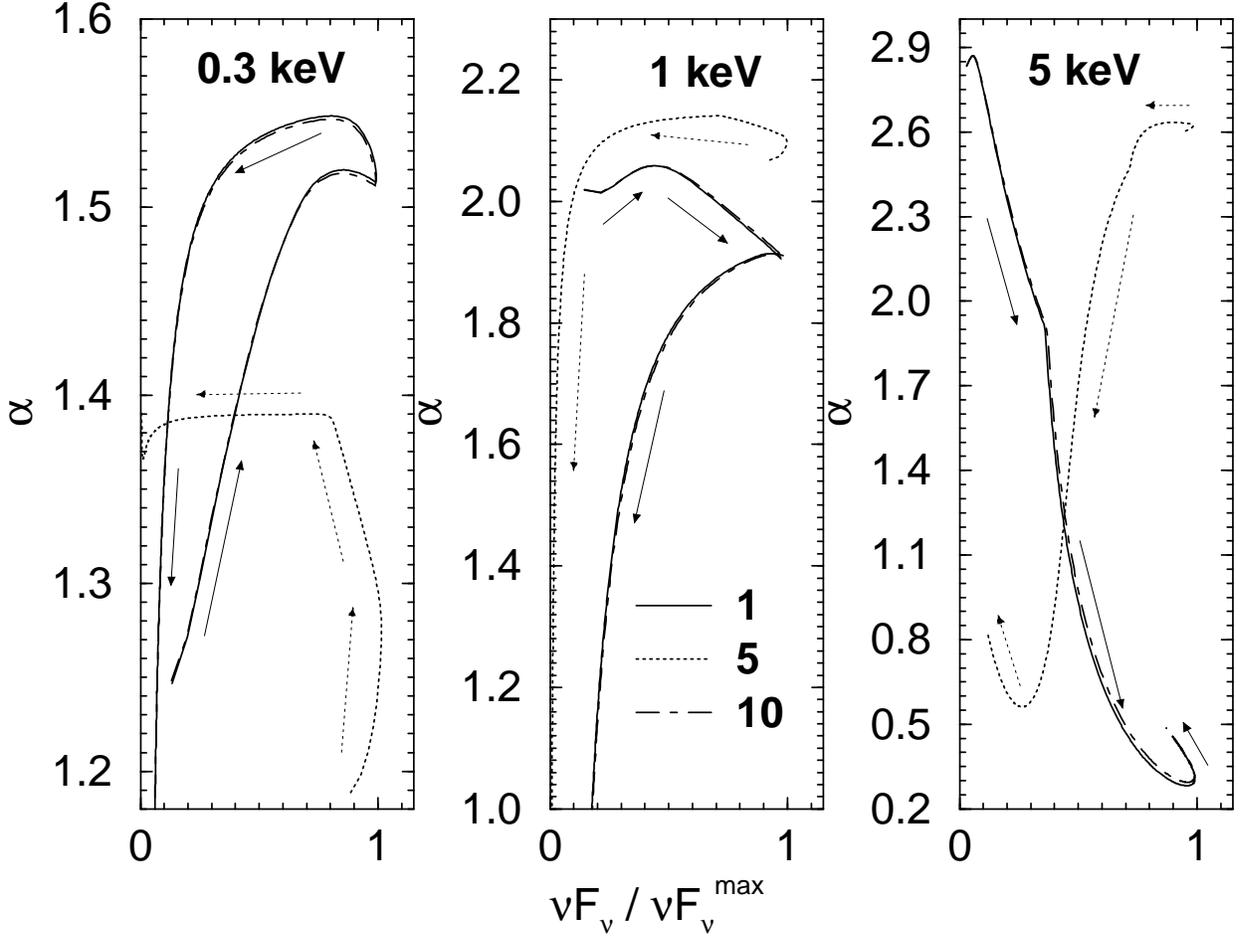}
\caption{Tracks in the hardness-intensity plane resulting from the time-dependent
simulations corresponding to spectral fits no. 1, 5, and 10, at 3 different X-ray
energies. While the SSC dominated model no. 5 produces drastically different
tracks in the hardness-intensity plane than the synchrotron dominated cases, 
there is no significant difference due to the addition of a moderate external-Compton 
component (fit no. 10). }
\label{hic}
\end{figure}

\newpage

\begin{figure}
\plotone{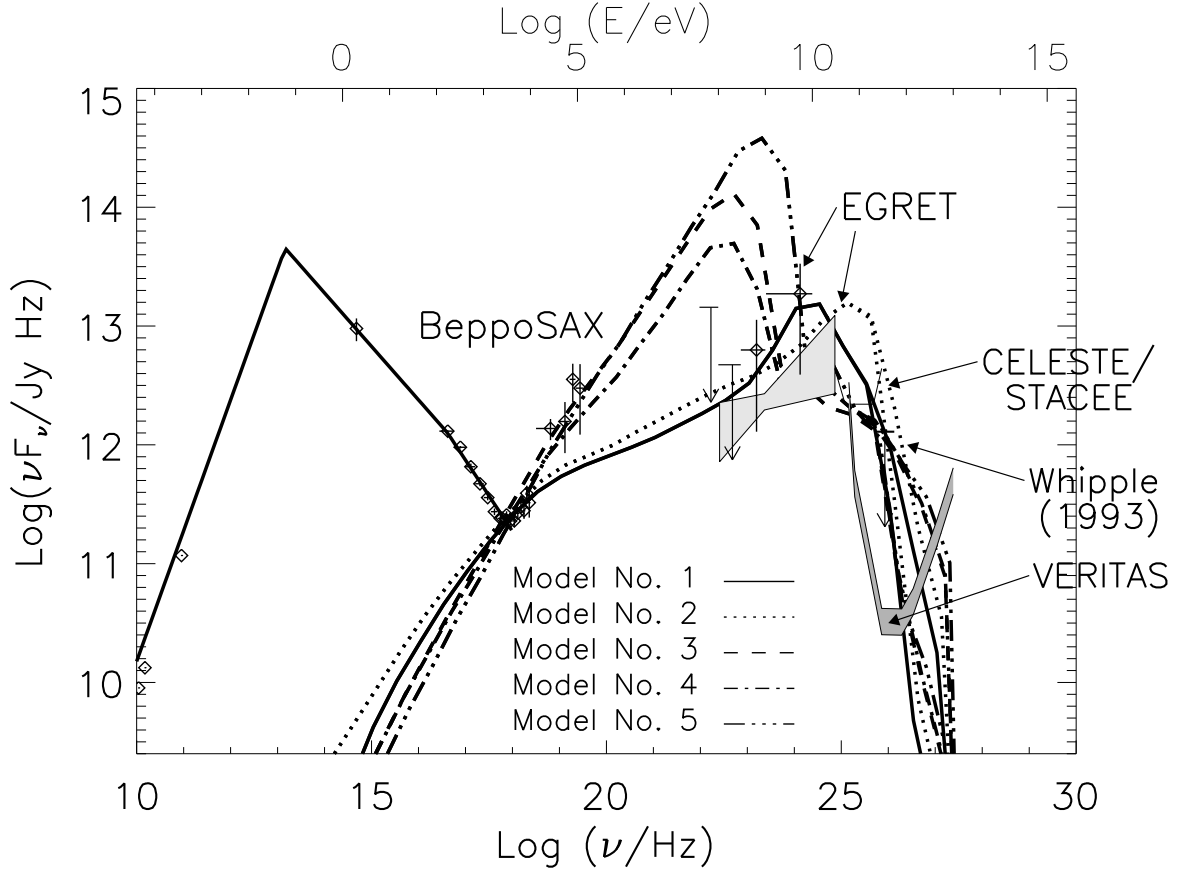}
\caption{Various model fits to the SED of W Comae in March/May 1998,
using the hadronic SPB model. All models are corrected for absorption in 
the cosmic background radiation field using the background models in 
\cite{aharonian01}. The two high-frequency branches of the model curves 
indicate the resulting fluxes using the two extreme background models in 
\cite{aharonian01}. The target photon field for $p-\gamma$ interactions 
is the primary electron synchrotron photon field, approximated by broken 
power laws with break energies $\epsilon_{b,1} = 0.06$~eV, $\epsilon_{b,2} 
= 173$~eV in the observer frame and photon spectral indices $\alpha_1 = 0.9$, 
$\alpha_2 = 2.45$ and $\alpha_3 = 2.6$. This target photon field is used 
as an input into the Monte-Carlo code to predict the high-energy component. 
For the complete list of model parameters, see Table \ref{SPBparameters}.}
\label{SPBfits}
\end{figure}

\newpage

\begin{figure}
\plotone{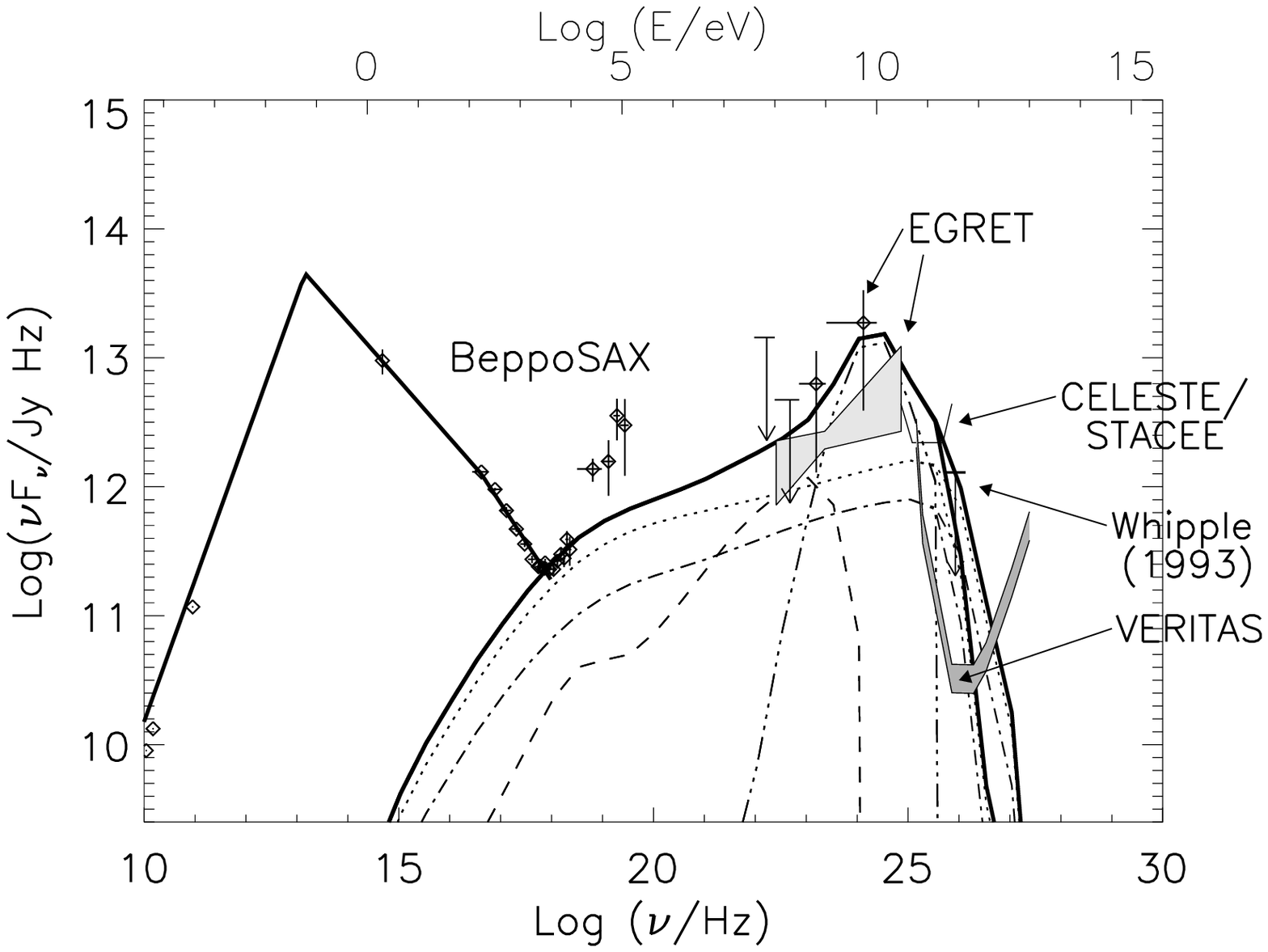}
\caption{Emerging cascade spectra for SPB model 1. $p$ synchrotron cascade 
(dashed line), $\mu$ synchrotron cascade (dashed-triple dot), $\pi^0$ cascade 
(dotted line) and $\pi^{\pm}$-cascade (dashed-dotted line), total (solid line). 
All model fluxes are corrected for absorption in the cosmic photon 
background as described in Fig. \ref{SPBfits}.}
\label{SPBfit1}
\end{figure}

\newpage

\begin{figure}
\plotone{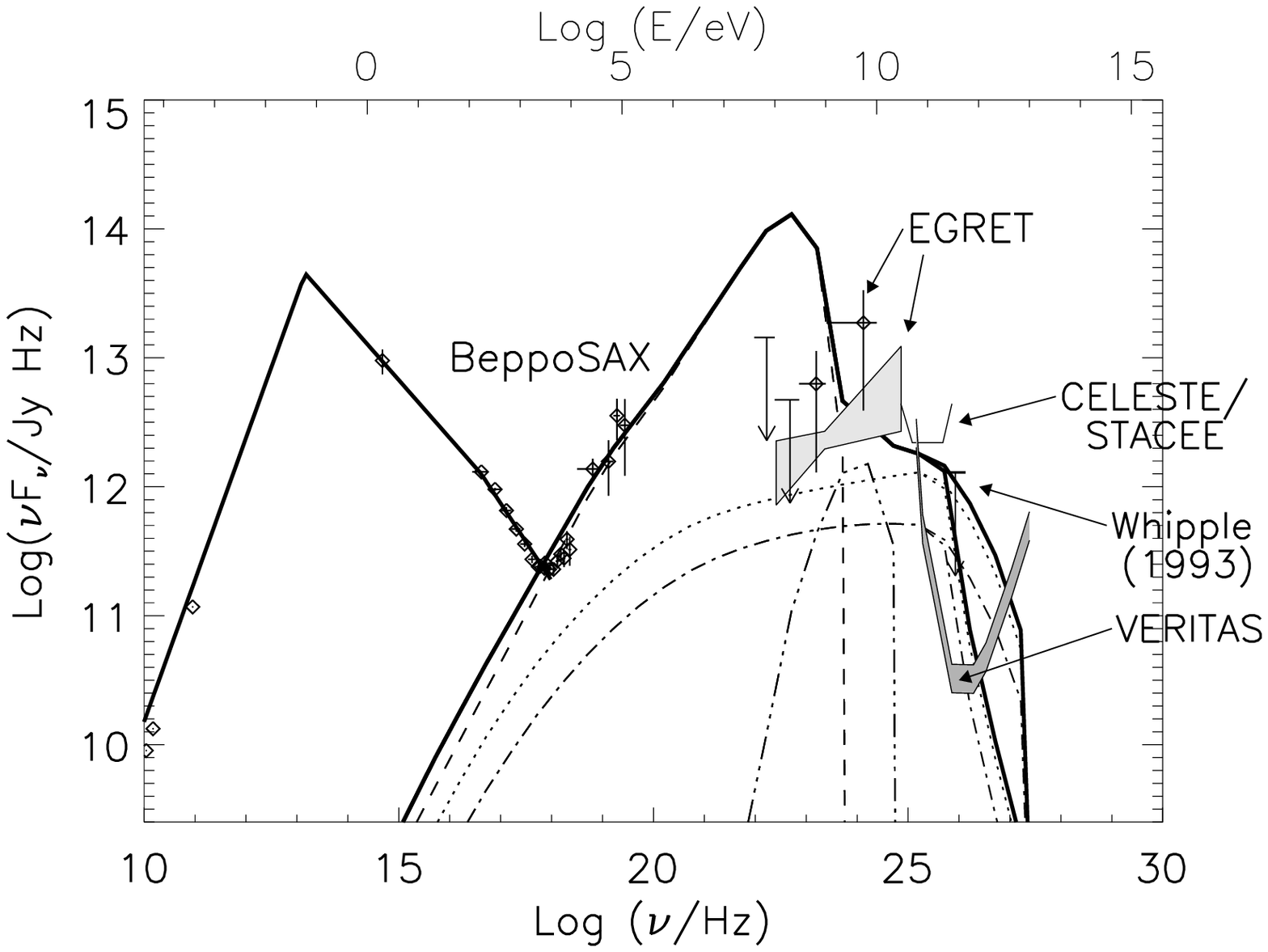}
\caption{Emerging cascade spectra for SPB model 3. $p$ synchrotron cascade 
(dashed line), $\mu$ synchrotron cascade (dashed-triple dot), $\pi^0$ cascade 
(dotted line) and $\pi^{\pm}$-cascade (dashed-dotted line), total (solid 
line). All model fluxes are corrected for absorption in the cosmic 
photon background as described in Fig. \ref{SPBfits}.}
\label{SPBfit3}
\end{figure}

\newpage

\begin{figure}
\plotone{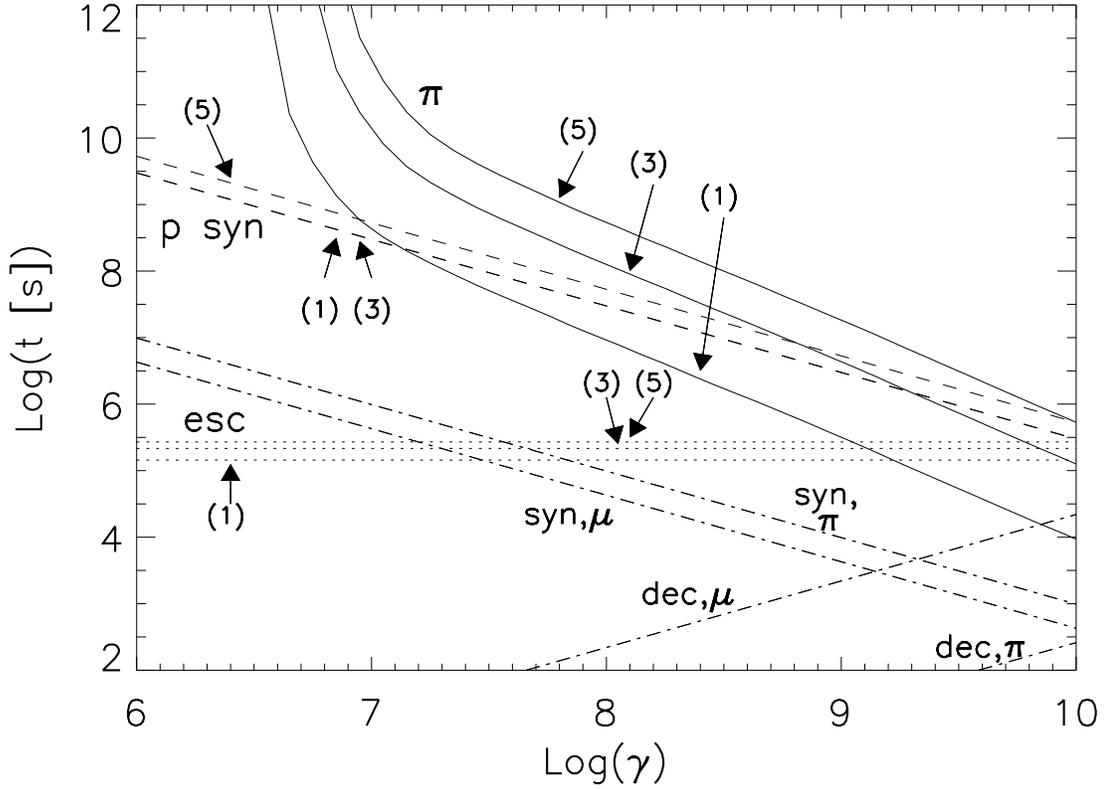}
\caption{Mean energy loss times (jet frame) for the hadronic models (1), (3) 
and (5) for
 $\pi$-photoproduction ($\pi$, solid lines), p synchrotron radiation 
(p syn, dashed lines) 
and escape (dotted lines). Loss times for $\pi^\pm$ and 
$\mu^\pm$ for synchrotron radiation (syn $\pi$, syn
 $\mu$) are also shown and 
compared with their mean decay time
 scales (dec $\pi$, dec $\mu$) for models 
(1) and (3).
}
\label{loss_timescales}
\end{figure}

\newpage

\begin{figure}
\plotone{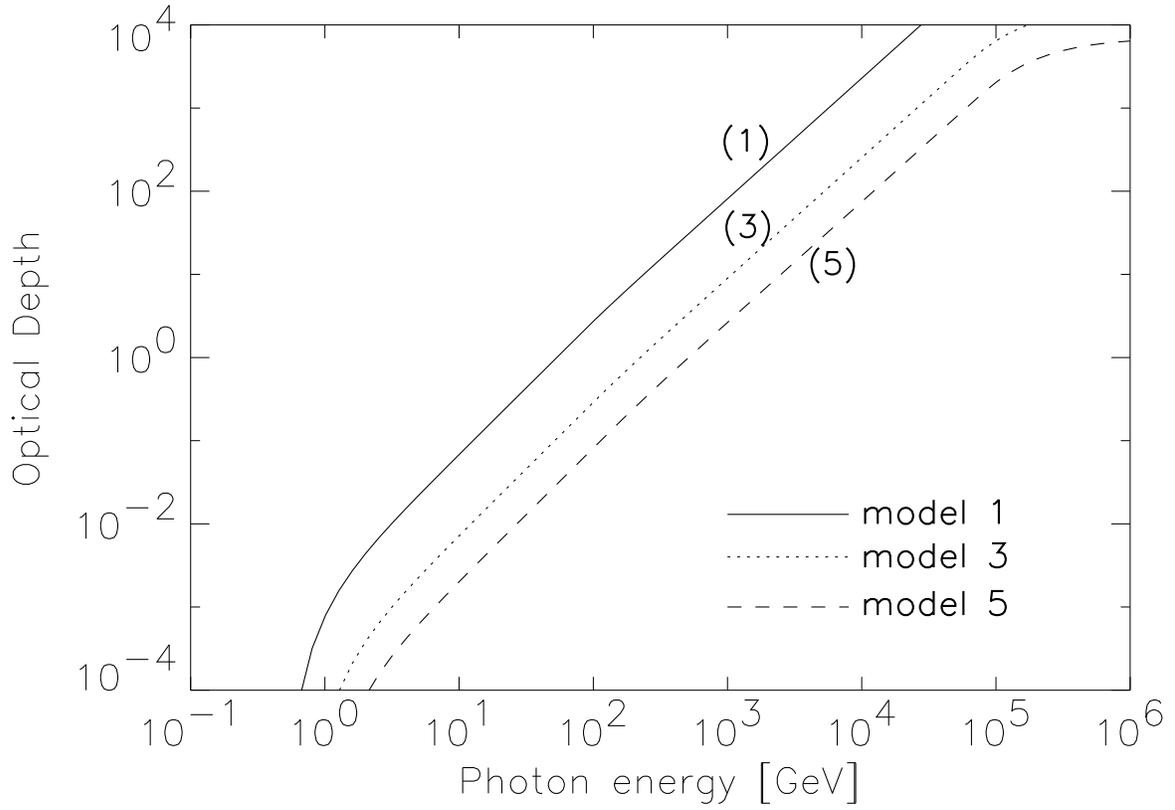}
\caption{Optical depth for photon absorption
 in the internal (primary 
electron synchrotron) target photon field for the
 hadronic models (1), (3) 
and (5). Photon energies refer to the co-moving frame of the jet.}
\label{tau_gg}
\end{figure}

\newpage

\begin{deluxetable}{cccccc}
\tabletypesize{\scriptsize}
\tablecaption{Log of EGRET observations of W~Comae. The flux in column 4 is the 
$> 100$~MeV flux in units of $10^{-8}$~photons~cm$^{-2}$~s$^{-1}$; the spectral
index $\alpha$ is the photon index. Upper limits are at the 2~$\sigma$ level.}
\tablewidth{0pt}
\tablehead{
\colhead{VP} & \colhead{Start date} & \colhead{End date} & \colhead{Flux} 
& \colhead{signif. [$\sigma$]} & \colhead{$\alpha$}
}
\startdata
0020  &  05/30/91 &  06/08/91  &  $36.4 \pm 13.2$ &    2.9 &  \\
0030  &  06/15/91 &  06/28/91  &         $< 20.2$ &    0.0 &  \\
0040  &  06/28/91 &  07/12/91  &   $10.8 \pm 3.9$ &    3.5 &  \\
0110  &  10/03/91 &  10/17/91  &         $< 25.0$ &    0.0 &  \\
204+  &  12/22/92 &  01/12/93  &  $23.4 \pm 10.7$ &    2.8 &  \\
2180  &  04/20/93 &  05/05/93  &  $13.7 \pm  6.9$ &    2.5 &  \\
2220  &  05/24/93 &  05/31/93  &         $< 35.0$ &    0.2 &  \\
304+  &  10/19/93 &  10/25/93  &  $15.4 \pm  4.7$ &    4.2 &  \\
3086  &  11/23/93 &  12/01/93  &  $28.1 \pm  9.4$ &    4.3 & $2.02 \pm 0.47$ \\
311+  &  12/13/93 &  12/20/93  &  $17.4 \pm  6.9$ &    3.1 &  \\
3130  &  12/27/93 &  01/03/94  &  $30.9 \pm 12.7$ &    3.1 &  \\
3220  &  04/05/94 &  04/19/94  &  $12.2 \pm  7.0$ &    2.2 &  \\
3260  &  05/10/94 &  05/17/94  &  $32.0 \pm 13.0$ &    3.6 &  \\
4060  &  12/13/94 &  12/20/94  &         $< 56.4$ &    0.2 &  \\
4180  &  04/25/95 &  05/09/95  &  $53.6 \pm 14.1$ &    5.3 & $1.88 \pm 0.29$ \\
5150  &  02/20/96 &  03/05/96  &  $19.2 \pm  7.1$ &    3.3 & $2.54 \pm 0.56$ \\
7155  &  03/20/98 &  03/27/98  &  $38.2 \pm 19.3$ &    2.7 & $1.27 \pm 0.58$ \\
9111  &  02/23/00 &  03/01/00  &        $< 107.2$ &    0.0 &  \\
\enddata
\label{history}
\end{deluxetable}

\newpage

\begin{deluxetable}{ccccccccc}
\tabletypesize{\scriptsize}
\tablecaption{Parameters of  the various fits shown in Figs. \ref{sscgraph} and \ref{blrgraph}. $\gamma_{1,2}$, and 
$p$ are the low- and high-energy cutoffs and the spectral index of the electron injection spectrum, $D$ is the Doppler
factor, $\tau_{\rm T, BLR}$ is the radial Thomson depth of the BLR. The magnetic field $B$ is the equipartition value.}
\tablewidth{0pt}
\tablehead{
\colhead{Fit no.} & \colhead{$\gamma_1$} & \colhead{$\gamma_2$} & \colhead{$p$} & \colhead{$n_e$ [cm$^{-3}$]} &
\colhead{$B$ [G]} & \colhead{$D$} & \colhead{$\tau_{\rm T, BLR}$} & \colhead{$F (> 40  \; {\rm GeV})$
[ph cm$^{-2}$ s$^{-1}$]}
}
\startdata
1 & 500 & $8.0 \times 10^4$ & 2.7 & 25 & 0.78 & 19.41 & 0 & $3.71 \times 10^{-11}$ \\
2 & 700 & $7.5 \times 10^4$ & 2.6 & 30 & 1.04 & 19.41 & 0 & $8.33 \times 10^{-11}$ \\
3 & 1000 & $6.2 \times 10^4$ & 2.5 & 35 & 1.37 & 19.41 & 0 & $1.62 \times 10^{-10}$ \\
4 & 1300 & $5.3 \times 10^4$ & 2.3 & 45 & 1.88 & 19.41 & 0 & $3.25 \times 10^{-10}$ \\
5 & 1600 & $4.8 \times 10^4$ & 2.2 & 55 & 2.33 & 19.41 & 0 & $4.45 \times 10^{-10}$ \\
6 & 1000 & $9.0 \times 10^4$ & 2.4 & 18 & 1.04 & 7.78 & 0 & $8.59 \times 10^{-11}$ \\
7 & 1000 & $4.0 \times 10^4$ & 2.2 & 65 & 2.43 & 28.05 & 0 & $6.83 \times 10^{-10}$ \\
8 & 500 & $8.0 \times 10^4$ & 2.7 & 25 & 0.78 & 19.41 & $1 \times 10^{-3}$ & $4.66 \times 10^{-11}$ \\
9 & 500 & $8.0 \times 10^4$ & 2.7 & 25 & 0.78 & 19.41 & $3 \times 10^{-3}$ & $8.32 \times 10^{-11}$ \\
10 & 500 & $8.0 \times 10^4$ & 2.7 & 25 & 0.78 & 19.41 & $1 \times 10^{-2}$ & $1.11 \times 10^{-10}$ \\
\enddata
\label{parameters}
\end{deluxetable}

\newpage

\begin{deluxetable}{cccccccccc}
\tabletypesize{\scriptsize}
\tablecaption{Parameters of the various fits shown in Figs. \ref{SPBfits}.
$n_p \propto \gamma_{p}^{-\alpha_p}$ is the injection proton spectrum in the 
energy range $2\leq\gamma_p\leq\gamma_{\rm{p,max}}$, $u_p$ is the injected proton 
energy density, $e/p$ the injected relativistic primary electron-to-proton number 
ratio, $D$ is the Doppler factor, $R_b$ the size of the emission region and 
$B$ the magnetic field. All parameters are measured in the jet frame. The 
predicted integrated photon fluxes above 40 GeV and 1 TeV, $F (> 40{\rm GeV})$ 
and $F(> 1{\rm TeV})$, are corrected for photon absorption in the cosmic
background radiation field using the different background models in \cite{aharonian01}.}
\tablewidth{0pt}
\tablehead{
\colhead{Fit no.} & \colhead{$\alpha_p$} & \colhead{$\gamma_{\rm{p,max}}$} & \colhead{$u_p$ [erg cm$^{-3}$]} & \colhead{$e/p$} &
\colhead{$D$} & \colhead{$R_b$ [$10^{15}$cm]} & \colhead{$B$ [G]} & \colhead{$F (> 40  \; {\rm GeV})$
[ph cm$^{-2}$ s$^{-1}$]} & \colhead{$F (> 1  \; {\rm TeV})$ [ph cm$^{-2}$ s$^{-1}$]}
}
\startdata
1 & 1.5 & $1 \times 10^9$ & 240 & 1.6 &  8 & 4.3 & 40 & $(4.5-6.2)\times 10^{-10}$ & $(0.7-11)\times 10^{-14}$\\
2 & 2.0 & $3 \times 10^9$ & 300 & 0.2 & 10 & 5.4 & 40 & $(8.3-13)\times 10^{-10}$  & $(1.5-8.9)\times 10^{-14}$\\
3 & 1.5 & $5 \times 10^8$ & 150 & 0.1 & 12 & 6.5 & 40 & $(0.8-1.3)\times 10^{-10}$ & $(2.6-5.4)\times 10^{-14}$\\
4 & 1.5 & $5 \times 10^8$ & 150 & 0.1 & 12 & 6.5 & 30 & $(0.9-1.4)\times 10^{-10}$ & $(2.8-5.9)\times 10^{-14}$\\
5 & 1.5 & $1 \times 10^9$ &  60 & 0.1 & 15 & 8.1 & 30 & $(0.7-1.0)\times 10^{-10}$ & $(1.8-48)\times 10^{-14}$\\
\enddata
\label{SPBparameters}
\end{deluxetable}

\end{document}